\documentclass[twocolumn,prb,aps]{revtex4}
\usepackage{amsmath}
\usepackage{amssymb}
\usepackage{graphicx}
\usepackage{dcolumn}
\usepackage{graphicx}
\usepackage{dcolumn}
\usepackage{bm}

\begin{document}

\preprint{Phys.Rev.B}

\title{Viscous transport and Hall viscosity in a two-dimensional electron system.}
\author{G. M. Gusev,$^1$ A. D. Levin,$^1$ E. V. Levinson, $^1$
 and A. K. Bakarov $^{2,3}$}

\affiliation{$^1$Instituto de F\'{\i}sica da Universidade de S\~ao
Paulo, 135960-170, S\~ao Paulo, SP, Brazil}
\affiliation{$^2$Institute of Semiconductor Physics, Novosibirsk,
630090, Russia}
\affiliation{$^3$Novosibirsk State University, Novosibirsk, 630090,
Russia}

\date{\today}
\begin{abstract}
Hall viscosity is a nondissipative response function describing momentum transport in two-dimensional (2D) systems with broken time-reversal symmetry. In the classical regime, Hall viscosity contributes to the viscous flow of 2D electrons in the presence of a magnetic field.
We observe a pronounced, negative Hall resistivity at low magnetic field in a mesoscopic size, two-dimensional electron system, which is attributed
to Hall viscosity in the inhomogeneous charge flow. Experimental results supported by a theoretical analysis confirm that the conditions for observation of Hall viscosity are correlated with predictions.


\end{abstract}

\maketitle

Considerable progress has been made recently in the non-perturbative understanding of the interaction effects in the electronic transport properties of metals within a hydrodynamic framework [1]. A hydrodynamic description is valid when the electron-electron scattering time is much shorter than the electron-impurity or electron-phonon scattering times.
The theory of the hydrodynamic regime, where transport is dominated by a viscous effect, has been developed in many theoretical studies [2-8]. It has been shown that shear viscosity contribution can be especially enhanced in the case where the mean free path due to electron-electron interaction $l_{ee}$ is much less than the sample width $W$, and the transport mean free path $l$ is in the order of or greater than the width : $l >> W$. In such a hydrodynamic regime, resistivity is proportional to the electron shear viscosity $\eta=\frac{1}{4}v_{F}^{2}\tau_{ee}$, where $v_{F}$ is the Fermi velocity and $\tau_{ee}$ is the
electron-electron scattering time $\tau_{ee}=l_{ee}/v_{F}$ [2]. It has been predicted that resistance decreases with the square of temperature,
$\rho \sim \eta \sim \tau_{ee} \sim T^{-2}$, and with the square of the sample width $\rho \sim W^{-2}$ [2-8].

The works demonstrating a feasible way to realize a hydrodynamic regime, so far, have been achieved in experiments with electrostatically defined GaAs wires [9,10] and graphene [11]. Until very recently,
experimental studies have been carried out in zero external magnetic field. In order to describe the large negative magnetoresistance in GaAs with high mobility electrons [13], the theoretical approach has been extended to include magnetohydrodynamic behaviour of two-dimensional systems [12]. Similar magnetoresistance has been observed in previous studies [14-16],
which could be interpreted as a manifestation of the viscosity effects. Recently it has been demonstrated that Palladium Cobaltate wires [17] and mesoscopic GaAs structures [18]
allow the study of underlying physical principles of the viscous system in a magnetic field and the carrying out of experiments to confirm theoretical predictions [12].

One interesting property  of a 2D fluid is Hall viscosity, which describes a nondissipative response function to an external magnetic field [13-29]. It is remarkable that, besides the importance of Hall viscosity in the context of condensed matter physics [19], it has been demonstrated that Hall viscosity arises in many different and seemingly
unconnected fields such as hydrodynamics, plasma, and liquid crystals [30]. It has been shown that classical Hall viscosity can be extracted from transport measurements in the emergent magneto-hydrodynamic regime in 2D electron systems [31-33]. Note that such a possibility has been questioned in a paper [12], where just the conventional Hall effect was found.
 However, one must take into account the higher order terms in the expansion of the electron distribution function by the angular harmonics of the electron velocity (related to inhomogeneities of a flow) [34].
Therefore the experimental study of the Hall resistivity in a viscous system may provide a useful platform for future theoretical developments in Hall viscosity.

\begin{figure}[ht]
\includegraphics[width=8cm]{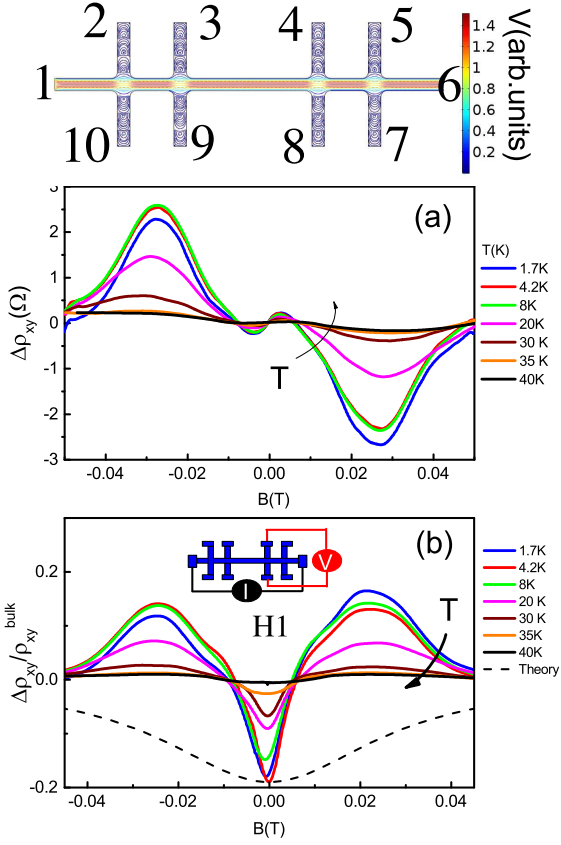}
\caption{(Color online) Top-sketch of the velocity profile for viscous flow in the experimental set up used in this study.
(a) Temperature dependent deviations from the conventional Hall resistivity $\Delta\rho_{xy}(T)$ of a mesoscopic GaAs well. (b) The ratio $\Delta\rho_{xy}(T)/\rho_{xy}^{bulk}$ for different temperatures. Dashes - theory with parameters described in the main text.}
\end{figure}

\begin{figure}[ht]
\includegraphics[width=8cm]{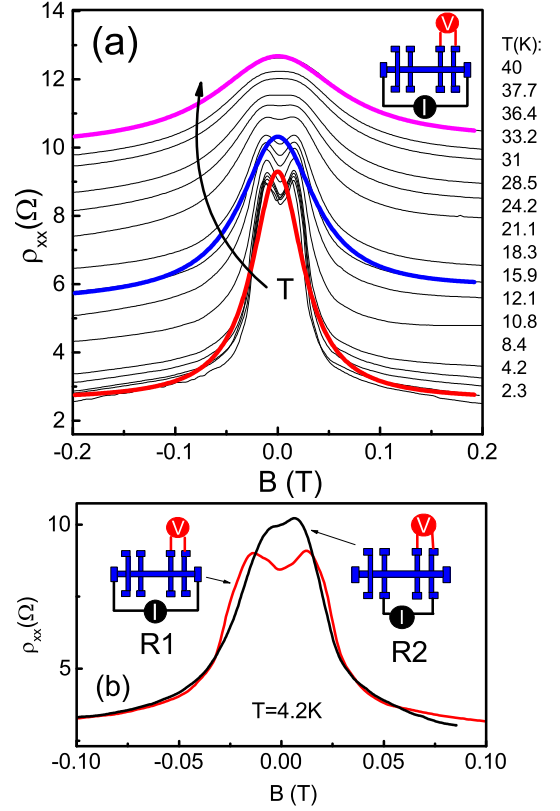}
\caption{(Color online)
(a) Temperature dependent magnetoresistance of a mesoscopic GaAs quantum well. Thick curves are examples illustrating magnetoresistance calculated from Eqs. (1,2). for  different temperatures: (a) 2.3 K (red), 21.1 K(blue) and 40 K( magenta); (b) 4.2 K (red), 19.2 K (blue) and 37.1 K ( magenta). (b) Comparison of the magnetoresistance for different configurations. The schematics show how the current source and the voltmeter are connected for measurements.}
\end{figure}

In the present study, we have gathered  all requirements for observation of the hydrodynamic effect and Hall viscosity in a 2D electron system and present experimental results accompanied by quantitative analysis. For this purpose, we chose GaAs mesoscopic samples with high mobility 2D electrons. We  employ commonly used longitudinal resistance, magnetoresistance, and the Hall effect to characterize electron shear viscosity, electron-electron scattering time, and reexamine electron transport over a certain temperature range, 1.5-40 K.
We  observe negative corrections to the Hall effect near zero magnetic field, which we attribute to classical Hall viscosity.

  Our samples are high-quality GaAs quantum wells with a width of 14~nm and electron density $n\simeq 9.1\times10^{11}$~cm$^{-2}$ at $T=1.4$~K.
 Parameters characterizing the electron system are given in Table 1.
The Hall bar is designed for multi-terminal measurements. The
sample consists of three $5 \mu m$ wide consecutive segments of
different length ($10, 20 , 10 \mu m$), and 8 voltage probes.
The measurements were carried out in a VTI cryostat, using a conventional
lock-in technique to measure the longitudinal $\rho_{xx}$ and  Hall $\rho_{xy}$  resistivities with an ac current of $0.1 - 1 \mu A$ through the sample, which is sufficiently low to avoid overheating effects. We also compare our results with the transport properties of 2D electrons in a macroscopic sample [34]. Three mesoscopic Hall bars from the same wafer were studied.
\begin{table}[ht]
\caption{\label{tab1} Parameters of the electron system in a mesoscopic sample at T=1.4 K. Parameters are defined in the text.}
\begin{ruledtabular}
\begin{tabular}{lccccccc}
$n_{s}$ & $\mu $ & $v_{F}$ & $E_{F}$ & $l$  & $l_{2}$ & $\eta$\\
$(cm^{-2}$) & $(cm^{2}/Vs$) & $(cm/s)$ & (meV) & $(\mu m$) & $(\mu m$) & $(m^{2}/s)$\\
\hline

$9.1\times10^{11}$ & $2.5\times10^{6}$ & $4.1\times10^{7}$ & $32.5$ & $40$ & $2.8$ & $0.3$\\
\end{tabular}
\end{ruledtabular}
\end{table}

Figure 1 shows deviations from conventional Hall resistivity $\Delta\rho_{xy}(T)=\rho_{xy}(T)- \rho_{xy}^{bulk}$ (referred to as H1 configuration) as a function of temperature. In order to determine
the bulk Hall resistivity $\rho_{xy}^{bulk}$, we measured the Hall effect in mesoscopic samples in a wider interval of the magnetic field ($-0.2T < B < 0.2 T$) and high $T\sim 40 K$ temperature. Indeed, we found $\rho_{xy}^{bulk}=-B/en_{s}$, where $e$ is the electron charge.  Figure 1(b) shows the ratio $\Delta\rho_{xy}(T)/\rho_{xy}^{bulk}$
for different temperatures. One can see a strong ($\sim 10-20\%$) deviation from the linear slope. The slope is opposite to the bulk Hall slope at low fields and has the same sign (negative for electrons) at large positive magnetic field and low temperatures.
Before analyzing the Hall effect quantitatively and in order to make this analysis more complete, we also measured longitudinal magnetoresistivity $\rho_{xx}(B)$ in the conventional configuration (referred to as R1).
Note, that the longitudinal magnetoresistance has been studied in
previous research for different configurations of the current and voltage probes [18]. Figure 2(a) shows $\rho_{xx}(B)$ as a function of magnetic field and temperature. One can see two characteristic features: a giant negative magnetoresistance $(\sim 400-600 \%)$ with a Lorentzian-like shape (except of the small feature near the zero field) and a pronounced temperature dependence of the zero field resistance.
\begin{figure}[ht]
\includegraphics[width=8cm]{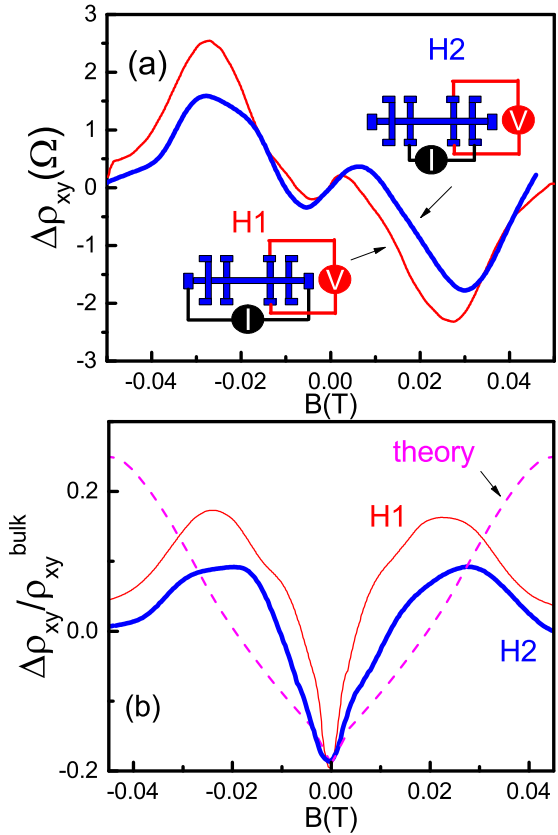}
\caption{(Color online)
(a) Hall effect for two configurations, $T=4.2 K$. (b) The ratio $\Delta\rho_{xy}(T)/\rho_{xy}^{bulk}$ for different configurations. Dashes (magenta) present calculations from ballisti + hydrodynamic theory with parameters described in the main text.}
\end{figure}
In general we expect that the character of the viscous flow strongly depends on the geometry and probe configurations [11]. Figure 2(b) shows a comparison of the  magnetoresistance measurements
in two configurations: conventional R1 configuration, and when the current is injected between probes 9 and 7 and the voltage is measured between probes 4 and 5 (referred to as R2 configuration).
Strikingly, the resistance at zero magnetic field increases in amplitude and the width of the Lorentzian magnetoresistance is slightly reduced.  The features near zero magnetic field are also smeared out. Surprisingly, we found that the resistance at B=0 is independent of temperature for the R2 configuration [35]. We attribute these results to the enhancement of the viscous contribution, and further, we prove it by a quantitative comparison with theory.
Furthermore we check the Hall resistance in a modified probe configuration [35]. Figure 3 shows a comparison of the Hall effect in H1 configuration with H2 configuration, where the current is injected between probes 9 and 7 and the voltage is measured between probes 4 and 8. One can see that $\Delta\rho_{xy}$ at low magnetic field is wider in H2 configuration, and, therefore, the ratio $\Delta\rho_{xy}(T)/\rho_{xy}^{bulk}$ exhibits a wider negative peak near zero B.

Classical transport can be characterized on different length scales: the Ohmic case $(l<<W)$, ballistic regime $(W<<l,l_{ee})$ and the hydrodynamic regime $(l_{ee}<<W<<l$).
\begin{figure}[ht]
\includegraphics[width=8cm]{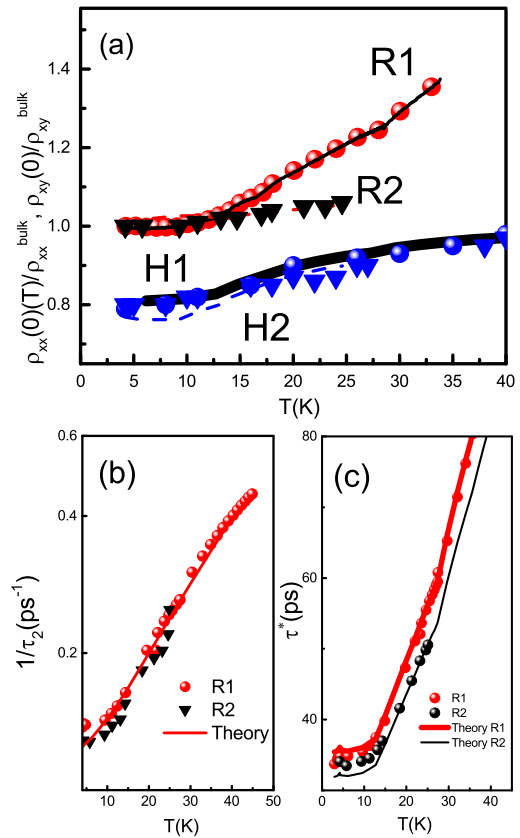}
\caption{(Color online)
(a) Temperature dependent resistivity and the Hall effect of a GaAs quantum well at $(B\rightarrow 0)$  for different configurations.
The solid lines and dashes  show calculations based on theoretical Eqs. (1) and (2)  with numerical parameters described in the main text. (b) Relaxation time $\tau_{2}$ as a function of temperature obtained by fitting the theory with experimental results. The solid line is the theory. (c) Relaxation time $\tau^{*}$ as a function of temperature. The solid lines are the theory with parameters presented in the main text.}
\end{figure}
In real samples, electrons are scattered by static defects, phonons, and the sample edge. All these processes can be expressed in terms of the scattering relaxation time $\tau$ and the boundary slip length $l_{s}$. Boundary no-slip conditions correspond to the ideal hydrodynamic case of diffusive boundaries with $l_{s}=0$, while the opposite limit (free surface boundary conditions) corresponds to the ideal ballistic case with $l_{s}=\infty$.

In the hydrodynamic approach, the semiclassical treatment of the electron transport describes the motion of carriers, when the higher order moments of the distribution function are taken into account. The momentum relaxation rate $1/\tau$ is determined by electron interaction with phonons and static defects (boundary).
The second moment relaxation rate $1/\tau_{2}$ leads to the viscosity and contains the contribution from electron-electron scattering and temperature independent scattering by disorder [12]. It has been shown that conductivity obeys the additive relation and is determined by two independent $\textit{parallel}$ channels: the first is due to momentum relaxation time and second due to viscosity [12,31].
This approach allows the introduction of the magnetic field dependent viscosity tensor and the derivation of the magnetoresisivity tensor [12,31-33]:
\begin{equation}
\rho_{xx}= \rho_{0}^{bulk}\left(1+\frac{\tau}{\tau^{*}}\frac{1}{1+(2\omega_{c}\tau_{2})^{2}}\right),\,\,\,
%3
\end{equation}
\begin{equation}
\rho_{xy}= \rho_{xy}^{bulk}\left(1-r_{H}\frac{2\tau_{2}}{\tau^{*}}\frac{1}{1+(2\omega_{c}\tau_{2})^{2}}\right),
%4
\end{equation}

where $\rho_{0}^{bulk}=m/ne^2\tau$, $\tau^{*}=\frac{W(W+6l_{s})}{12\eta}$, viscosity $\eta=\frac{1}{4}v_{F}^{2}\tau_{2}$, $r_{H}$ is numerical coefficient in the order of 1 [12].
At the limit of zero magnetic field $(B\rightarrow 0)$, one obtains negative corrections to Hall resistivity due to Hall viscosity in the limit of small $l_{s}$:  $\rho_{xy}= \rho_{xy}^{bulk}\left[1-6r_{H}(\l_{2}/W)^{2}\right]$.

It is instructive to collect the equations for relaxation rates separately:
$\frac{1}{\tau_{2}(T)}=A_{ee}^{FL}\frac{T^{2}}{[ln(E_{F}/T)]^{2}}+\frac{1}{\tau_{2,0}}$, and $\frac{1}{\tau(T)}=A_{ph}T+\frac{1}{\tau_{0}}$, where $E_{F}$ is the Fermi energy, the coefficient $A_{ee}^{FL}$ can be expressed via the Landau interaction parameter, however, it is difficult to calculate quantitatively (see discussion in [12]).
The  term $A_{ph}$ is due to scattering electrons by acoustic phonons [36,37], and $\frac{1}{\tau_{0}}$ is the scattering rate due to static disorder.
Note that effective relaxation time $\tau^{*}$ is proportional to the rate $\frac{1}{\tau_{2}}$ (not time).
We represent the evolution of  $\rho_{xx}$ at B=0  with temperature
in Fig. 4(a) for configurations R1, R2.
We fit the magnetoresistance curves in Fig. 2 and the resistance in zero magnetic field shown in Fig. 4(a) with the 3 fitting parameters : $\tau(T)$, $\tau^{*}(T)$ and $\tau_{2}(T)$. Comparing the temperature dependencies, we extract the following parameters:
$\tau_{2,0}=0.8\times10^{-11}$ s, $A_{ee}^{FL}=0.9\times10^{9} s^{-1}K^{-2}$, $l_{s}=3.2 \mu m$, $A_{ph}=10^{9}s^{-1}K^{-1}$, $\tau_{0}=5\times10^{-10}s$ for configuration R1. For configuration R2 all parameters are the same, except of $l_{s}=2.8 \mu m$.
Assuming that the viscous effect is small in a macroscopic sample,
we attempt to reduce the number of independent parameters by measuring $\rho_{0}(T)\sim 1/\tau(T)$  and extracting $A_{ph}$ independently [35].
However, we find a parameter in the macroscopic sample $A_{ph}^{macr}=1.3\times10^{9} s^{-1}K^{-1}$, which is slightly higher than in the mesoscopic sample [35].
Table 1 shows the mean free paths : $l=v_{F}\tau$, $l_{2}=v_{F}\tau_{2}$  and viscosity, calculated with parameters, which we extracted from the fit with experimental data.
Fig. 4(b) shows the dependencies of $1/\tau_{2}(T)$ and $\tau^{*}(T)$ extracted from the comparison with theory.
Note, that $\tau^{*}(T)$ depends on the boundary conditions, and the difference in its behaviour for configurations R1 and R2 could be explained by the difference in the parameter $l_{s}$. More diffusive boundary conditions (smaller value of $l_{s}$) correspond to stronger hydrodynamic effects.

Now we return to the issue of Hall viscosity.
Fig. 3(b) shows the dependence $\rho_{xy}/\rho_{xy}^{bulk}$ at $B\rightarrow 0$ as a function of temperature for configurations H1 and H2 with calculations obtained independently from magnetoresistance
measurements. From comparison with the experiment, we find the adjustable parameter $r_{H}=0.4$. This value agrees with numerical calculations performed in the model [31], where parameter $r_{H}\approx0.35$ was obtained. The existence of the parameter $r_{H}<1$ simply reflects the fact that the viscous Hall correction in Eq. (2) never exceeds $100\%$, which one expects even for a small ratio $\l_{2}/W$ (see, for example,  $l_{2}/L=0.04$ and $W/l=0.1$, considered in Fig. 2(b) of the ref. [31]).

Figure 1(b) shows the Hall curve as a function of B calculated from Eq. (2). Note that the theory predicts a broad Loreantzian-like peak, while rapid change of the sign is observed near $B\approx 0.01 T$.
The discrepancy could be related to the higher order expansion terms of the angular velocity harmonics of the electron velocity, which are not considered for longitudinal
magnetoresistivity [12].

 It is important to note that, in the ballistic regime, $\rho_{xx}$ and $\rho_{xy}$ strongly depend on the magnetic field due to the size effects [38-41]. Unfortunately the changing B-scale is almost the same $\sim W/R_{L}$ ($R_{L}=mV_{F}/eB$ is the Larmor radius)
for both contributions [31], and ballistic and hydrodynamic effects can obscure each other. The magnitude of the ballistic contribution depends on the ratio $W/l$. In addition the relative ballistic contribution  $\rho_{xx}^{ball}/\rho_{0}^{bulk}$ exhibits strong variation with $ W/R_{L}$ because the resistivity directly depends on the relaxation time $\tau$ through the boundary scattering, while relative contribution to the Hall effect  $\rho_{xy}^{ball}/\rho_{xy}^{bulk}$ is almost independent of $ W/R_{L}$, since the Hall effect does not depend on the relaxation time (but rather the size effect)[37-39].
  Note, that the sign of the effects is the same: the ballistic contribution leads to an increase in boundary scattering, an increase of $\rho_{xx}$, amplification of the classical Hall slope at $W/R_{L}=0.55$, and quenching of the Hall effect near B=0 [39,40].
From comparison with theory, at low temperatures, we found that $\rho_{xx}^{ball} < \rho_{0}$ (see Fig. 2(a)).
 We attempted to fit the magnetoresistance curves with a smaller Lorentzian amplitude, considering the features near $W/R_{L}=0.55$ due to the ballistic contribution, and found the fitting parameters $\frac{\tau}{\tau^{*}}$ only $10\%$ smaller. Note also, that since the
ballistic and hydrodynamic contributions have the same sign, the B-scale of the magnetoresistance is almost the same, when $\rho_{xx}^{ball}$ is added to magnetoresistance.
However, for the same parameters, $\rho_{xy}^{ball}$ is comparable with the hydrodynamic contribution and the ballistic corrections tend
to counteract the hydrodynamic corrections in the Hall effect. The ballistic model predicts the quenching of $\rho_{xx}^{ball}$ near B=0 [40,41], therefore, $\rho_{xy}/\rho_{xy}^{bulk}$ is not affected by the ballistic effect in very close proximity of zero field. However, the ballistic contribution leads to a decrease in the B-scale of the  $\rho_{xy}(B)$, when $\rho_{xx}^{ball}$ is added to the Hall effect.
We performed calculation of the ballistic transport in our sample geometry [35]. We confirmed that the billiard model reproduces earlier numerical calculations.
  Figure 3(b) shows our numerical results together with the hydrodynamic model.  Indeed ballistic contribution results in a decrease of the width of the negative peak near B=0.
One can see that, for H2 configuration with stronger hydrodynamic effects (smaller $l_{s}$), the calculated curve could be brought in better agreement with measurements, indicating the relevance of this explanation.

In conclusion, we have measured the evolution of longitudinal and  Hall resistivities with temperature in high quality GaAs quantum wells.  Our observations are correlated  with the predictions of classical Hall viscosity for electron flow.

We thank P.S.Alekseev and Z.D.Kvon for helpful discussions. The financial support of this work by FAPESP, CNPq (Brazilian agencies) is acknowledged.

\section{Supplementary matterial: Viscous transport and Hall viscosity in a two-dimensional electron system.}

\subsection{Dependence of  magnetoresistance and Hall effect on temperature in a mesoscopic Hall bar sample}

In general we expect that the character of the viscous flow strongly depends on the geometry and probe configurations of the sample. For example, the velocity field in a curved pipe flow
becomes more inhomogeneous than in a straight pipe, which may enhance the viscosity effect. In the main text, we present the results for two configurations: conventional R1 configuration and R2 configuration, where the current is injected between probes 9 and 7, and the voltage is measured between probes 4 and 5.
Figure 5(a) shows $\rho_{xx}(B)$ as a function of magnetic field and temperature for R2 configuration. Similarly to configuration R1, one can see a giant negative magnetoresistance $(\sim 400-600 \%)$ with a Lorentzian-like shape. However, the magnetoresistance at zero magnetic field is weakly dependent on the temperature.
We fit the magnetoresistance curves in Fig. 1(a) and the resistance in zero magnetic field shown in Fig. 4(a) of the main text with 3 fitting parameters : $\tau(T)$, $\tau^{*}(T)$ and $\tau_{2}(T)$. Comparing the temperature dependencies, we extract the following parameters: $\tau_{2,0}=0.8\times10^{-11}$ s, $A_{ee}^{FL}=0.9\times10^{9} s^{-1}K^{-2}$, $l_{s}=2.8\mu m$, $A_{ph}=10^{9}s^{-1}K^{-1}$, $\tau_{0}=5\times10^{-10}$ s.
 Note, that $\tau^{*}(T)$ depends on the  boundary conditions, and the difference in the behavior between configurations R1 and R2 could be explained by a difference in parameter $l_{s}$. More diffusive boundary  conditions (smaller value of $l_{s}$) correspond to stronger hydrodynamic effects.
 Figure 5(b) shows deviations from conventional Hall resistivity $\Delta\rho_{xy}(T)=\rho_{xy}(T)- \rho_{xy}^{bulk}$ in H2 configuration as a function of temperature.
 We compare these results with those from configuration H1 in the main text.
 \begin{figure}[ht]
\includegraphics[width=8cm]{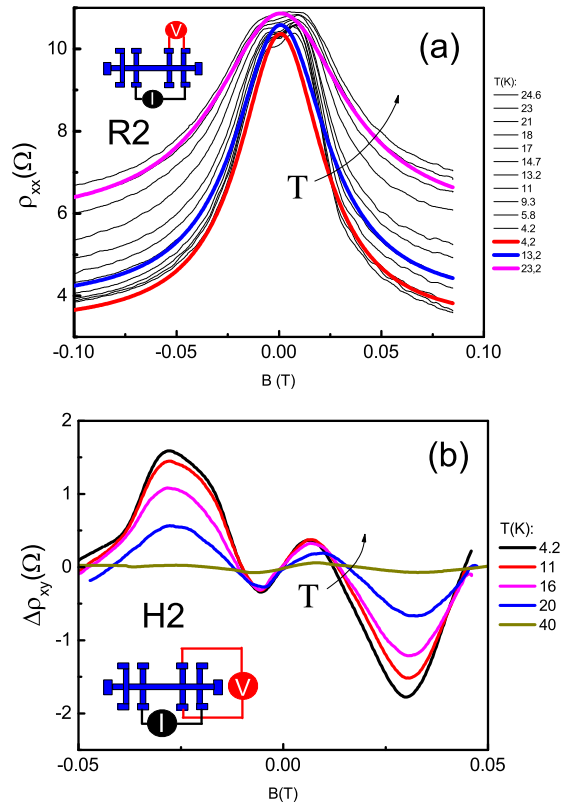}
\caption{(Color online)
(a) Temperature dependent magnetoresistance of a mesoscopic GaAs quantum well for R2 configuration. Thick curves
are examples illustrating magnetoresistance calculated from Eq. 1. for  different temperatures: (a) 4.2 K (red), 13.2 K(blue) and 23.2 K( magenta).
(b) Temperature dependent deviations from the conventional Hall resistivity $\Delta\rho_{xy}(T)$ of a mesoscopic GaAs well
for H2 configuration. The schematics in inserts show how the current source and the voltmeter are connected for the measurements. }
\end{figure}

\subsection{Dependence of  magnetoresistance on temperature in a macroscopic Hall bar sample}

According to model [1], the viscosity contribution is represented by an extra relaxation mechanism which contains the relaxation time
of the second moment of the electron distribution function due to electron-electron collisions. The conventional bulk momentum relaxation mechanism and  time $\tau(T)$  are related to
scattering by acoustic phonons and static disorder. Therefore, it is expected that in a wide microscopic sample the viscosity effects are small, and we can extract the dependence of $\tau(T)$ by measuring the resistivity in such a sample. After that, the dependence $\tau(T)$  can be used to determine $\tau^{*}(T)$. With this aim, we measure the macroscopic sample. The   samples   have   Hall-bar   geometry   (length $l\times$ width $W = 500 \mu m \times 200 \mu m$) with  six  contacts. Figure 6(a) shows the longitudinal magnetoresistivity $\rho_{xx}$ measured in the local
configuration for a macroscopic  Hall bar sample as a function of magnetic field and temperature. One can see a linear growth of the zero filed resistivity $\rho_{0}$ with increasing T (Fig. 6(b)).
We still observe residual hydrodynamic effects, which confirms the importance of e-e interactions in 2D transport.
Note that theoretical paper [1] presents an attempt to compare
experimental magnetoresistance obtained in a previous study in a wide macroscopic GaAs sample [2] with theory. Although the theory can reproduce the shape of magnetoresistance and its evolution
with temperature, the value of the magnetoresistance and T-dependence of $\rho_{0}(T)$ can be reproduced only by the introduction of some effective width $W_{eff}<<W$.
Assuming the viscous time $\tau^{*}$ is much larger than $\tau$, we are able to fit a linear dependence with phonon+static disorder scatterer  with  adjustable parameters $A_{ph}^{macr}=1.3\times10^{9} s^{-1}K^{-1}$ and $\tau_{0}=9.1\times10^{-11} s$.
For comparison we represent the result for a mesoscopic sample, described in the man text, with R1 configuration.
The parameters are slightly different : $A_{ph}=10^{9}s^{-1}K^{-1}$, $\tau_{0}=5\times10^{-10}$ s. Note also that at low temperatures
the mesoscopic sample reveals a deviation from  linear dependence. In general it would be expected that $\tau_{0}$ in a mesoscopic sample saturates in the ballistic regime when $l\sim L,W $.
Previous measurements [3] obtained the value $A_{ph}=0.8\times 10^{9}s^{-1}K^{-1}$ ($n_{s}=6\times cm^{-2}$), which grows linearly with $n_{s}$ and is found in good agreement with the theory for the deformation potential constant $D=13.5eV$ [4].

\begin{figure}[ht]
\includegraphics[width=8cm]{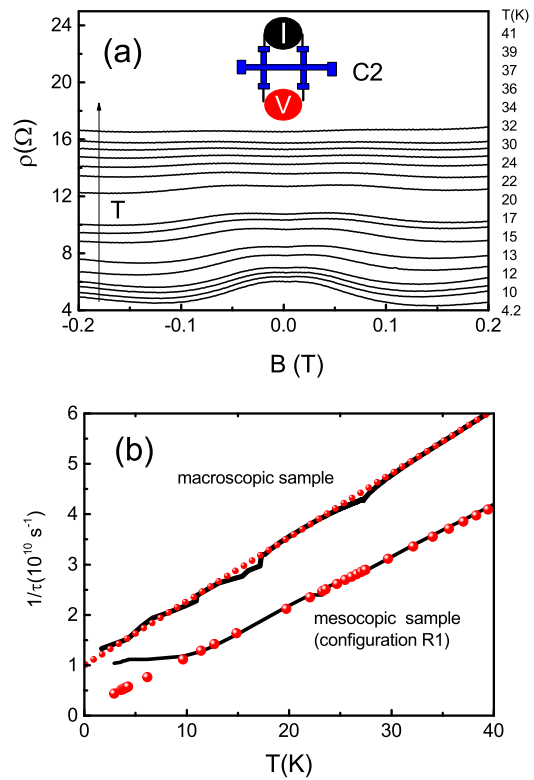}
\caption{(Color online)
(a) Temperature dependent magnetoresistance of a GaAs quantum well in a macroscopic Hall bar sample.  The schematics show how the current source and the voltmeter are connected for the measurements.
(b) Solid lines show the inverse relaxation time $1/\tau$  for macroscopic and mesoscopic samples as a function of temperature obtained by fitting the theory with experimental results (Fig. 1(a)). The circles are theory with parameters presented in the text.}
\end{figure}
 In addition we measure the Hall effect in the macroscopic sample. We don`t find any deviation of the Hall effect from the linear dependence near a small magnetic field.

\subsection{Billiard model for longitudinal and Hall resistances}
In the main text, we describe the results of the measurements of longitudinal and Hall resistances in a mesoscopic sample and compare the results with the hydrodynamic model. In the ballistic case, the transport is dominated by scattering at the boundaries. In this section we describe the ballistic model based
on Landauer-Buttiker formalism. The resistivities and conductivities follow from a set of equations:\

\begin{figure}[ht]
\includegraphics[width=8cm]{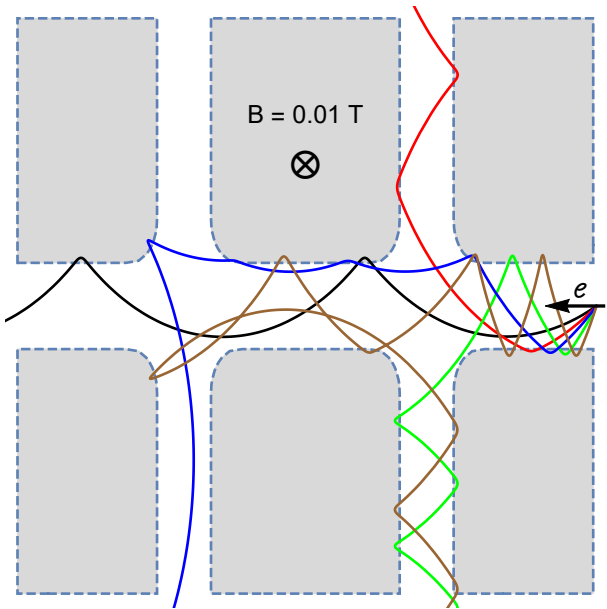}
\caption{(Color online)
Examples of electron trajectories inside Hall bar presented for five different injection angles. Electrons are injected from right lead with Fermi velocity $v_F=4.1 \times 10^5 m/s$, magnetic field is maintained fixed at $B=0.01 T$. Gray areas are Hall bar walls modeled as repulsive parabolic potential for electrons.}
\end{figure}

\[I_{i}\boldsymbol{=}\frac{e}{h}[(1-R_{ii})\mu_i-\sum_{j{\neq}i}T_{ij}\mu_j]\]

where $T_{ij}$ and $R_{ii}$ are the probabilities for carriers incident in the lead $j$ to be reflected or transmitted into lead $i$. Solving the system of these equations, we calculate resistance from injected current $I_i$ and measured potentials $\mu_j=eV_j$ using Ohm's law $ R=(V_n-V_m)/I$.

We simulate the injection of the $\sim10^4 $ numbers of the electrons towards different junctions. Examples of possible trajectories are presented in Fig. 7.

In our model the shape of the wall potential is considered to be parabolic. We estimate the steepness of the potential from the
assumption that the width of the region where the potential
increases from the bottom to the Fermi energy is
of the same order as the Fermi wavelength for typical electron
concentrations. We would like to emphasize that in our experiments we used samples with high electron density corresponding to the steeper potential.
 Assuming confinement edge potential $U = kx^2/2$ (for coordinates outside the Hall bar geometry), we estimate $k=0.008meV/ \buildrel _{\circ} \over {\mathrm{A}}^2$.
The results for Hall resistivity are used in the main text. We add the ballistic Hall contribution to the experimental result. The ballistic contribution leads to a decrease of the B-scale of $\Delta\rho_{xy}$ and brings the theoretical curve to a better agreement with experimental data.

\end{document}